\documentclass[10pt]{article}
\usepackage{amssymb,latexsym}
\usepackage{amsmath,amsbsy}
\usepackage{epsfig}

\DeclareMathOperator{\koneperp}{\mathbf{k}^{\perp}_{1}}
\DeclareMathOperator{\ktwoperp}{\mathbf{k}^{\perp}_{2}}
\DeclareMathOperator{\pperp}{\mathbf{p}^{\perp}}
\DeclareMathOperator{\Pperp}{\mathbf{P}^{\perp}}
\DeclareMathOperator{\koneplus}{k^{+}_{1}}
\DeclareMathOperator{\ktwoplus}{k^{+}_{2}}
\DeclareMathOperator{\Pplus}{P^{+}}
\DeclareMathOperator{\pplus}{p^{+}}
\DeclareMathOperator{\qperp}{\mathbf{q}^{\perp}}
\DeclareMathOperator{\pmag}{p^{\perp}}
\DeclareMathOperator{\kperp}{k^{\perp}}
\DeclareMathOperator{\kplus}{k^{+}}

\begin{document}

\title{\hskip7cm NT@UW-00-30\\
  Trouble in Asymptopia---the Hulth\'en Model on the Light Front}
\author{ B. C. Tiburzi and G. A. Miller\\
        Department of Physics\\ 
University of Washington\\      
Box $351560$\\  
Seattle, WA $98195-1560$ }
\date{\today}
\maketitle

\begin{abstract}
We use light-front dynamics to calculate the electromagnetic
form-factor for the Hulth\'en model of the deuteron. For small momentum 
transfer $Q^2 < 5$ GeV${}^2$ the relativistic effects are quite small.
For $Q^2 \sim 11$ GeV${}^2$
 there is $\sim 13 \%$ discrepancy between the relativistic and
 non-relativistic approaches. For asymptotically large
momentum transfer, 
however, the light-front form factor, $\sim \log{Q^2}/Q^4$,  markedly differs from the non-relativistic
version, $\sim 1/Q^4$.
This behavior is also present for any wave function, such as those obtained
from
realistic potential models, which can be represented as a sum of Yukawa functions.
Furthermore,  the asymptotic behavior is in disagreement with the 
Drell-Yan-West relation. We investigate precisely how to determine the asymptotic 
behavior and confront the problem underlying troublesome form factors on the
light front.    
\end{abstract}

\section{Introduction}

The light front approach to quantum dynamics was introduced
by Dirac \cite{D} a half century ago. 
Since then, light front dynamics has developed into an active area of research for a variety of reasons,
e.g. its minimal set of dynamical operators, the
simplicity of the light-front vacuum, and the close connection to experimental
observables.
 Light front
techniques have long been used in analyzing high energy experiments with
nuclear and nucleon targets \cite{A, SJ, Y, Z}. 
Indeed light front dynamics is relevant to describe such reactions, since, for example: in the parton model,
the ratio $\kplus/\pplus$ (where $\kplus = k^{0} + k^3$ is the plus momentum of the struck quark and 
$\pplus$ that of the target) is nothing more than the Bjorken $x$-variable. 

Some recent efforts have been made \cite{BBM} to render the theory more
understandable   by using 
models reminiscent of basic quantum mechanics rather than by invoking quantum field theory. These models 
find particular reality in nuclear physics \cite{M}, where some nucleon interactions may be described by a 
mean field potential. Nonetheless, the similarity of the light-front bound state equation to the Schr\"odinger 
equation is grounds enough to put familiar quantum mechanical problems on the light front. Below, we do precisely 
this for the Hulth\'en model of the deuteron and its electromagnetic form factor. Of particular concern here is the asymptotic
behavior of the form factor, which differs from the non-relativistic
version. This may be of interest to
experimentalists
seeking to probe asymptopia. Recent measurements of deuteron form factors at the Jefferson National Laboratory 
\cite{J} reach momentum transfers of $Q^2 = 6$ GeV${}^2$ and future projects hope to reach upwards of $Q^2 = 11$ GeV${}^2$. 
In this range of momentum transfer, there is $\sim 13 \%$ discrepancy between the relativistic and non-relativistic 
form factors calculated in this paper (as we will illustrate in figure \ref{f:exp}).

This paper's organization is similar to that of a detective story. First we recall a minimal amount of light front 
dynamics in section \ref{s:lfd} and explain how we put the non-relativistic Hulth\'en potential on the light front. 
Next in section \ref{s:ff}, we calculate the electromagnetic form factor using light front dynamics and compare with 
the non-relativistic version calculated in section \ref{s:nrff}. The low momentum behavior of the these form factors 
shows only minimal differences, while the high momentum behavior leads to surprising trouble in asymptopia (section 
\ref{s:asym}). We could solve the mystery at this point by deriving the asymptotic behavior of the form factor. 
Instead, we proceed by assuming factorization holds in the asymptotic limit. This leads us to consider various previous 
attempts at dealing with the end-point region and to dispel any lingering misconceptions. In section \ref{s:sim}, we 
discover that troublesome asymptotic behavior also lurks in other models on the light front. With enough clues 
at hand, we are able to pin-point the cause. The asymptotic behavior is then
deduced in section \ref{s:ca}, and is similar to that obtained for the
Wick-Cutkosky model in Ref.~\cite{V2}.
Finally, 
we summarize our findings in a brief concluding section.

\section{Hulth\'en model on the light front}
  \label{s:lfd}
In light front dynamics, one quantizes the fields at equal light-front time specified by 
$x^{+} = x^0 + x^3 = t + z$. This redefinition of the time variable leaves us with
a new spatial variable $x^{-} = x^0 - x^3 = t - z$. The remaining spatial variables 
are left unchanged by this transformation: $\mathbf{x}^{\perp} = (x^1, x^2)$. 

If one uses $x^{-}$ as a spatial variable, then its momentum conjugate is
$\pplus = p^0 + p^3$. This leaves $p^{-} = p^0 - p^3$ as the energy, or the 
$x^{+}$-development operator. The details of this formalism do not concern us here---the 
interested reader should consult \cite{H} for a good overview. 
What is important to note, however, is that 
the relativistic dispersion relation $p^{\mu}p_{\mu} = m^2$ takes the form
\begin{equation}
p^{-} = \frac{\pperp^2 + m^2}{\pplus},
\end{equation}
and thus the expression for the kinetic energy avoids the historically problematic square root. 

For a bound state of two particles interacting via a potential $V$, the 
light-front wave function is determined by solving the equation \cite{BL}
\begin{equation} \label{e:lfwf}
        \psi = \frac{1}{M^2 - \sum_{i = 1,2} \frac{k^{\perp}_{i}{}^2 + 
m_{i}^{2}}{x_{i}}} V \psi, 
\end{equation}
where $M$ is the invariant mass of the system, $m_{i}$ the particle mass, and  
$x_{i}$ is the plus momentum fraction carried by the $i^{\text{th}}$ particle, 
namely $ x_{i} = k^{+}_{i}/ \Pplus $, with $\Pplus$ as the total plus
momentum, $\koneplus + \ktwoplus$. Let us take
the nucleons to be of equal mass, and  use $m = \frac{m_{p} + 
m_{n}}{2}$ as the nucleon mass. Furthermore, since we have only two particles,
the sum of $x_{1}$ and $x_{2}$ is one. So we choose $x_{1} \equiv x$ and 
consequently, $x_{2} = 1 - x$.

In order to simplify Eq. \ref{e:lfwf}, it is customary to define the relative
light-front variables \cite{OT}
\begin{align}
\Pperp &=  \koneperp + \ktwoperp\\ \notag       
\pperp &=   - x \koneperp + (1-x) \ktwoperp.
\end{align} 
Straightforward algebra transforms Eq. \ref{e:lfwf} into
\begin{equation} \label{e:lfwf2}
        M^2 \psi = \Big( \frac{\pperp^2 + m^2}{x ( 1 - x)}  + V \Big) \psi, 
\end{equation}
which is  the coordinate representation of the 
Weinberg equation \cite{W}. Eq. \ref{e:lfwf2} is still quite complicated to
solve,
so we define an auxiliary operator 
\begin{equation} \label{e:p3}
p^3 = \Big( x - \frac{1}{2} \Big) \sqrt{ \frac{\pperp^2 + m^2}{x (1 - x)}}
\end{equation}
to cast the equation into a familiar form. Defining $M = 2 m - \epsilon$ 
(where $\epsilon$ is the  binding energy) and using the above 
definition, we find  
\begin{align} \label{e:schr}    
\Big( \frac{\epsilon^2}{4} - \epsilon m \Big) \psi & = \Big( \pperp^2 + (p^3)^2 + 
        \frac{V}{4}  \Big) \psi  \notag \\
        & \equiv \big( \mathbf{p}^2 + V^{H} \big) \psi, 
\end{align}
where we have efficaciously chosen $V^{H} = V / 4 $ to be the Hulth\'en 
potential \cite{h123}.

The above equation is the coordinate-space version considered by others, 
see e.g. \cite{Fu}. Taking $\bf{p}$ conjugate to $\bf{r}$, we have 
\begin{equation} \label{e:VH}
V^{H} ( \mathbf{r} ) = \frac{b^2 - a^2}{1 - e^{(b - a) r}},
\end{equation}
and the well known ground state solution
\begin{equation} \label{e:wfr}
        \psi(r) \propto \frac{e^{-a r} - e^{- b r}}{r}
\end{equation}
with $a = \sqrt{\epsilon m - \epsilon^2 / 4}$ as dictated by Eq.     
\ref{e:schr}.
The experimentally determined  values of the model parameters are
 \cite{CWW}: $a = 
0.23161 \; \text{fm}^{-1}$ and $ b = 1.3802 \; \text{fm}^{-1}$. 

\section{Electromagnetic Form Factor} \label{s:ff}
The electromagnetic form factor on the light front has the form \cite{GBB}
\begin{equation} \label{e:ff}
F(Q^2) = \int \frac{dx \; d\pperp}{x (1 - x)} \psi^{\text{*}}(x, \pperp + (1-x)
 \qperp) \psi(x, \pperp),
\end{equation}
where the momentum transfer $Q^2 = \qperp^2$. The momentum-space Hulth\'en 
wave function is the Fourier transform of our solution Eq. \ref{e:wfr}, 
namely 
\begin{align} \label{e:wfp}
\psi(x, \pperp) & \equiv  \frac{m \sqrt{N}}{4}\Bigg( \frac{1}{a^2 + \pperp^2 + 
(p^3)^2} - \frac{1}{b^2 + \pperp^2 + (p^3)^2}  \Bigg) \notag \\
& = \frac{m \sqrt{N} x (1-x)}{4 x (1-x) \alpha^2 + (2 x - 
1)^2 m^2 + \pperp^2} \Big( \delta^{\alpha}_{a} - \delta^{\alpha}_{b} \Big). 
\end{align}

To calculate the form factor, we must perform three integrals. Writing 
$d\pperp = \pmag d\pmag d\phi$ with $\phi$ as the angle between $\pperp$ and $\qperp$, 
we see the $\phi$ integral and subsequently the 
$\pmag$ integral can be computed analytically. Performing these integrals leaves 
us with
\begin{equation} \label{e:hff}
        F(Q^2) = \int_{0} ^{1} f \Big(x, (1-x)^2 Q^2 \Big) dx, 
\end{equation}
where 
\begin{equation} \label{e:f}
        f(x, k^2) =  m^2 N x (1-x) \; g_{\alpha \beta}(x,k^2) \Big( \delta^{\alpha}_{a}\delta^{\beta}_{a} - \delta^{\alpha}_{a}
\delta^{\beta}_{b} - \delta^{\alpha}_{b} \delta^{\beta}_{a} + \delta^{\alpha}_{b}\delta^{\beta}_{b}  \Big)
\end{equation}
with
\begin{equation}
g_{\alpha\beta}(x, k^2) = \frac{\pi}{\rho_{\alpha \beta}(x, k^2)}  \log \Bigg[     
\frac{ k^4  + k^2 \big( 2 \; \gamma_{\alpha}(x) + \gamma_{\beta}(x) + \rho_{\alpha\beta}(x, k^2)  \big) + 
\gamma_{\beta}(x) \big( \rho_{\alpha \beta}(x, k^2) - \gamma_{\beta}(x) - \gamma_{\alpha}(x) \big) }
{ \gamma_{\alpha}(x) \big( -k^2 +  \rho_{\alpha\beta}(x, k^2) + \gamma_{\beta}(x) - \gamma_{\alpha}(x) \big) } \Bigg] \notag
\end{equation}
where we have defined 
\begin{align} 
        \gamma_{\mu}(x) &\equiv 4 x (1-x) \mu^2 + (2x-1)^2 m^2 \notag \\
        \rho_{\alpha\beta}(x, k^2) & \equiv
        \sqrt{ k^4  + 2 k^2 \gamma_{\beta}(x) + \gamma_{\alpha}(x)^2 }. \notag 
\end{align}

\begin{figure}
\begin{center}
\epsfig{file=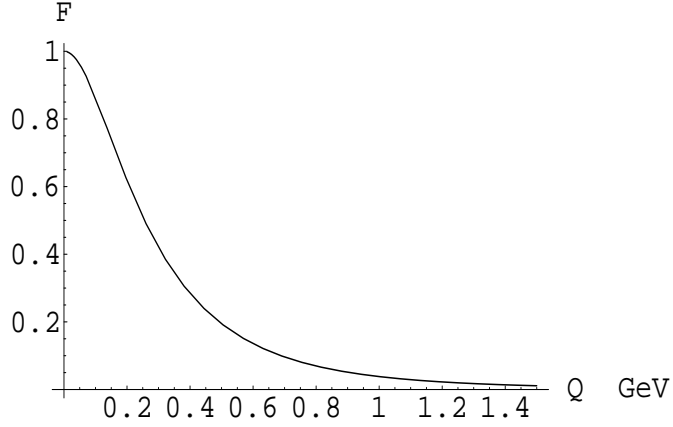,height=2.5in,width=3.5in}
\caption{Plot of the light front Hulth\'en form factor Eq. 
\ref{e:hff} as a function of $Q$ in GeV.}
\label{f:hff}
\end{center}
\end{figure}

We choose $N$ so that the form factor is normalized, $F(0) = 1$. The constant
$N$ is determined by setting  
 $Q= 0$ in Eq. \ref{e:ff}, which yields $N = 14.931^{-1}$.  Figure \ref{f:hff} shows the form 
factor as a function of $Q$. 
We have also calculated the derivative of $F(Q^2)$ in the limit $Q \to 0$ in 
order to find the root-mean-square deuteron radius
\begin{equation}
 \label{e:rms}
R_{rms} = \lim_{Q^2 \to 0} \sqrt{- 6 \frac{dF}{dQ^2}} = 1.9467 \; 
\text{fm}.
\end{equation}

\section{The Non-Relativistic Limit and Comparison} \label{s:nrff}

Our solution to the Hulth\'en model on the light front closely resembles the 
non-relativistic treatment. In fact, we have used the non-relativistic solution 
as a guide in constructing the relativistic wave function. Clearly relativistic 
effects are contained in the light-front variable $x$. Quite simply then,  
the non-relativistic limit of Eq. \ref{e:ff} is found in the limit $m \to 
\infty$ by retaining terms to $\mathcal{O}[\frac{1}{m}]$. Inverting Eq. \ref{e:p3} yields
\begin{equation}
        x =  \frac{1}{2} + \frac{p^3}{2 \sqrt{\pperp^2 + (p^3)^2 + m^2}} \approx 
\frac{1}{2} + \frac{p^3}{2 m}. 
\end{equation}
Since the measure $dx \; d\pperp \to d\mathbf{p}/2m$ is already first order, we need 
only keep leading order terms in the wave functions to find the non-relativistic form 
factor.  It is clear that to leading order: $\psi(x, \pperp) \to \psi(\mathbf{p})$,
 where the latter is the non-relativistic wave function. Quite similarly, we see
\begin{equation}
\psi(x, \pperp + (1-x) \qperp) \to \frac{1}{\alpha^2 + \mathbf{p}^2 + \pperp \cdot \qperp + \qperp^2/4} \big( \delta^{\alpha}_{a} 
- \delta^{\alpha}_{b}\big) \notag. 
\end{equation}
Returning to the expression for the form factor, the non-relativistic limit is then
\begin{equation}
F(Q^2) \to  \int d\mathbf{p} \; \psi^{*}(\mathbf{p} + \qperp/2) \psi(\mathbf{p})  .
\end{equation}
 The form factor above depends neither on the orientation of
 $\qperp$, nor of $\mathbf{p}$. Let us then rotate our coordinate system so that $\qperp$ is no longer
 completely transverse. This three-dimensional rotation is only possible now
 because 
 we are integrating $d\mathbf{p}$,  which can not be done in the light-front
 version.
 Thus we have sent $\qperp \to \mathbf{q}$ while maintaining the length, $\mathbf{q}^2 = Q^2$. After
 this rotation, the form factor is strikingly non-relativistic and can be computed 
analytically using Eq. \ref{e:wfr}
\begin{align} \label{e:hnr}
F(Q^{2})^{\text{NR}} & = \int d\mathbf{r} \; \big| \psi(\mathbf{r}) \big|^{2} e^{- i \mathbf{q} \cdot \mathbf{r} /2} \notag \\
 & = \frac{m N^{\prime}}{Q} \Bigg[ \tan^{-1} \Big( \frac{Q}{4 a} \Big) - 2 \tan^{-1} \Big( \frac{Q}{2 (a + b)} \Big) + \tan^{-1} \Big( 
\frac{Q}{4 b} \Big) \Bigg],  
\end{align}
with $N^{\prime}$ chosen to make $F(0)^{\text{NR}} = 1$. 

\begin{figure}
\begin{center}
\epsfig{file=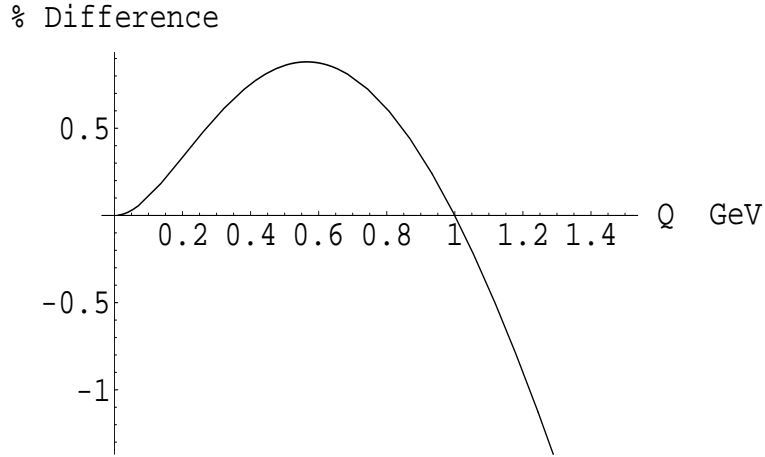,height=3in,width=4in}
\caption{Relativistic and non-relativistic form factor comparison: percent difference:
$100 \times (F^{\text{NR}} - F)/F^{\text{NR}}$ plotted as a function of $Q$ in GeV.}
\label{f:rnr}
\end{center}
\end{figure} 

From this analytical result, the rms. radius can be easily calculated (see Eq. \ref{e:rms})
\begin{equation}
R_{\text{rms}} = \frac{\sqrt{a^4 + 5 a^3 b + 12 a^2 b^2 + 5 a b^3 + b^4}}{2^{3/2} a b( a + b )} = 1.9395 \; \text{fm}.
\end{equation}
Comparing with our previous result, the relativistic system is larger by only $0.37 \%$. This
 confirms our suspicion that relativistic effects in this deuteron model are small. We can further
 confirm this by looking at the difference of the relativistic form factor Eq. \ref{e:hff} 
and the non-relativistic version Eq. \ref{e:hnr}. The percent difference is plotted for 
low $Q$ in Figure \ref{f:rnr} illustrating a difference only $\sim 1 \%$ in this momentum r\'egime. 
The small nature of relativistic effects was noted early on \cite{fs}.

\section{Trouble in  Asymptopia} \label{s:asym}
The above plot
shows the absolute percent difference continually increasing as $Q$ increases. In this section,
 we  investigate how the light-front form factor compares with the non-relativistic version for large $Q$.

\subsection{Exploring Asymptopia}
Given our analytic expression for the non-relativistic form factor Eq. \ref{e:hnr}, it is simple
 to Taylor expand about $Q = \infty$ to find
its  asymptotic behavior. To leading order
\begin{equation} \label{e:nrasym}
\lim_{Q \to \infty} F(Q^2) \sim \frac{64 a b (a + b)^2}{Q^4} = \frac{0.080585 \; (\text{GeV})^4}{Q^4}.
\end{equation}

The asymptotic behavior of the relativistic form factor is found with the aid of the Drell-Yan-West relation \cite{DYW} 
(under the assumption that the end-point region dominates the form factor for large $Q$). This relation takes the form
\begin{equation} \label{e:dyw}
\lim_{x \to 1} \;  x \; f(x, 0) \sim (x-1)^{2 \delta - 1} \Longleftrightarrow \lim_{Q \to \infty} F(Q^2) \sim (Q^2)^{-\delta}.
\end{equation}
The $x$-distribution function $f(x, 0)$ can be calculated analytically using Eq. \ref{e:ff} with $\qperp = 0$ 
and subsequently expanded about $x =1$. The leading-order term in the expansion is $\mathcal{O}[(x-1)^3]$, 
from which we deduce $1/Q^{4}$ behavior in asymptopia. Given that there were only small differences between 
the relativistic and non-relativistic form factors for low $Q$, we might expect agreement in the 
asymptotic region. Moreover, both form factors go as $1/Q^{4}$ for large $Q$. So when scaled by
 $Q^{4}$, at worst the form factors will tend to some common difference as $Q \to \infty$. 

\begin{figure}
\begin{center}
\epsfig{file=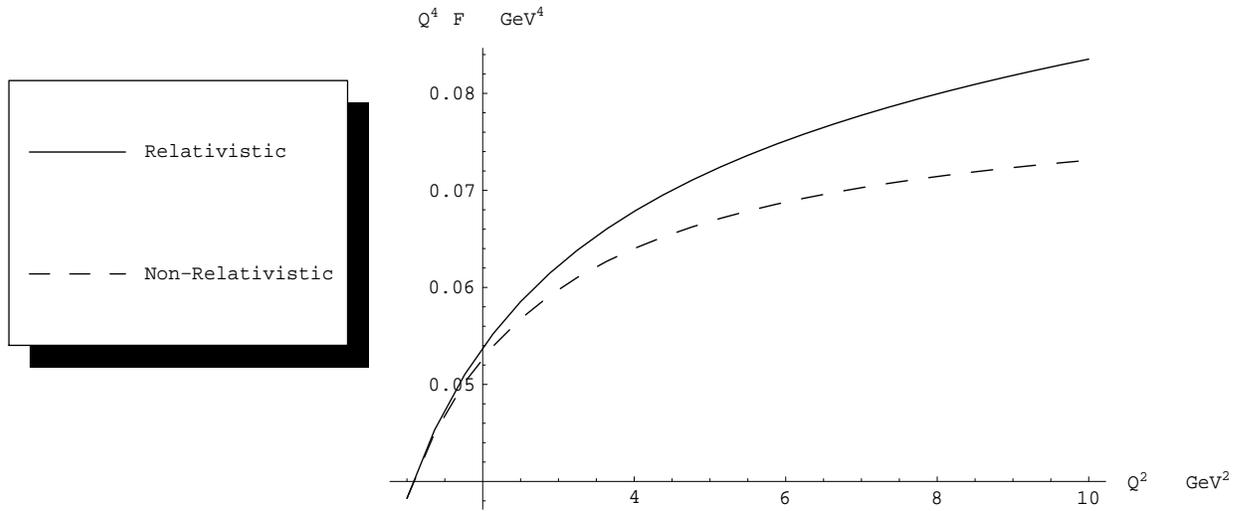,height=3in,width=6.75in}
\caption{Relativistic and non-relativistic comparison for experimentally relevant $Q^2$: 
the form factors are scaled by $Q^4$ and plotted as a function of $Q^2$ in GeV${}^2$.}
\label{f:exp}
\end{center}
\end{figure} 

\begin{figure}
\begin{center}
\epsfig{file=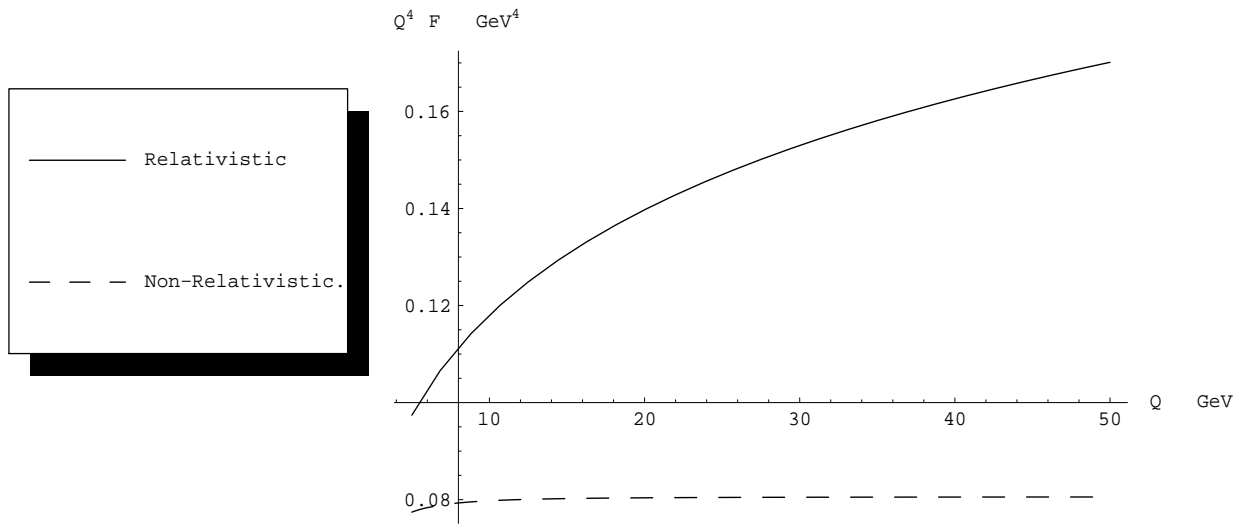,height=3in,width=6.75in}
\caption{Relativistic and non-relativistic comparison in the asymptotic limit: 
the form factors are scaled by $Q^4$ and plotted as a function of $Q$ in GeV.}
\label{f:rnrhighQ}
\end{center}
\end{figure} 

To compare the asymptotic behavior, we have plotted the relativistic and non-relativistic 
form factors (scaled by $Q^4$) for large $Q$ (in figure \ref{f:exp} we plot for experimentally relevant $Q^2$, whereas in figure 
\ref{f:rnrhighQ} we are mathematically contrasting the asymptotics). The non-relativistic form factor lines up 
well with the asymptotic limit predicted by Eq. \ref{e:nrasym}. The relativistic form 
factor, however, markedly differs from its non-relativistic counterpart, in disagreement
 with the Drell-Yan-West relation Eq. \ref{e:dyw}. We remind the reader that the relativistic form
factor is computed exactly  for our model \cite{numb}.
The huge disparity between the non-relativistic and relativistic results, shown
in  figure 
\ref{f:rnrhighQ}
warrants a complete journey through asymptopia.   

\subsection{Sharp Peaks at $x=1$}
\begin{figure}
\begin{center}
\epsfig{file=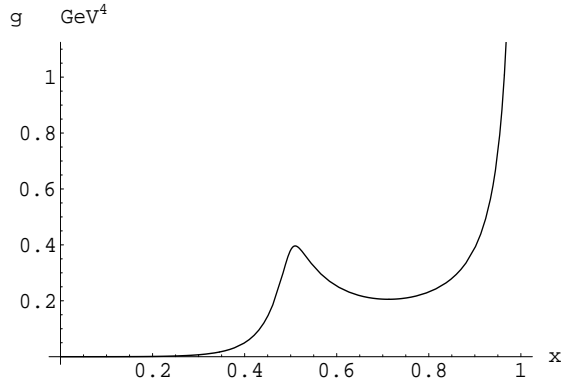,height=2in,width=3in}
\caption{Plot of $g(x)$ appearing in Eq. \ref{e:asymf2}. The singularity at $x=1$ is too severe to bound a 
finite area under the curve.}
\label{f:asymf}
\end{center}
\end{figure}

Before proceeding, we note that our light front wave function Eq. \ref{e:wfp} is properly behaved:
\begin{align}  \label{e:wellbehaved}
\pperp^2 \psi (x, \pperp) \to 0 &  \: \; \text{as} \: \pperp^2 \to \infty \: \: \;  \text{and} \notag \\
\psi (x, \pperp ) \to 0 & \: \; \text{as} \: x \to 0,1 
\end{align}
These conditions stem from non-relativistic versions, and will be trivially satisfied for light-front wave functions
created using Eq. \ref{e:p3} and \ref{e:schr}.   Thus knowing the 
non-relativistic wave functions are peaked for small momenta, our light-front Hulth\'en 
wave function must be peaked for small transverse momenta. For large momentum transfer 
$Q$, Eq. \ref{e:ff} shows that large momentum flows through either $\psi$ or $\psi^{*}$.
 Following Brodsky and Lepage \cite{BL}, the dominant contributions to the form factor
 in the asymptotic limit come from the two regions which minimize wave function suppression
\begin{align} \label{e:regions}
& \text{i}) \: \: |\pperp| \ll |(1-x) \qperp|, & \text{where} \: \psi^{*}(\pperp + (1-x)\qperp) \: \text{is small and} \: \psi(\pperp) \: 
\text{is large} \notag \\
& \text{ii}) \:\:  |\pperp + (1-x)\qperp | \ll |(1-x)\qperp|, & \text{where} \: \psi(\pperp) \: \text{is small and} \: \psi^{*}(\pperp + 
(1-x) \qperp) \: \text{is large}  
\end{align}  
Working first in region i, we can neglect $\pperp$ relative to $\qperp$ in $\psi^{*}$ since 
the light-front wave functions are peaked for low transverse momenta. The contribution from
 region i is exactly the same as from ii which is made obvious by shifting $\pperp$.
 Thus dominant contributions to the form factor in the asymptotic r\'egime appear as
\begin{align} \label{e:asymf}
F(Q^2) & \approx 2 \int \frac{dx}{x (1-x)} \psi^{*}(x, (1-x) \qperp) \int d\pperp \; \psi(x, \pperp) \\
\label{e:asymf2}
& \approx \frac{8 \pi N m^2 (b^2 - a^2)}{Q^4} \int_{0}^{1} \frac{dx \; x^2}{(1-x)^2} \log \Bigg[ \frac{4 x (1-x) b^2 + 
(2x-1)^2 m^2}{4 x(1-x) a^2 + (2x-1)^2 m^2} \Bigg] 
\equiv \int_{0}^{1} \frac{dx \; g(x)}{Q^4},
\end{align}
where we have retained the same normalization constant that appears in Eq. \ref{e:f}.
\footnote{We obtained the same result by brute force
Taylor expansion of Eq. \ref{e:f} about $Q = \infty$. Since the integral over $x$ diverges, 
the series expansion of $f(x, Q^2)$ 
lacks uniform convergence.} But to determine the asymptotic behavior, 
we must perform the integral over $x$ which diverges! The end-point 
region is too peaked for the integral to converge, as illustrated by Figure \ref{f:asymf}.

The end-point region appears to dominate the form factor in asymptopia (as suggested in \cite{F}). Perhaps the actual 
asymptotic behavior can be extracted by putting the end-point region under
scrutiny. Above, we merely
assumed the 
validity of the Drell-Yan-West relation; now, we will rigorously investigate it for our model. 

\begin{figure}
\begin{center}
\epsfig{file=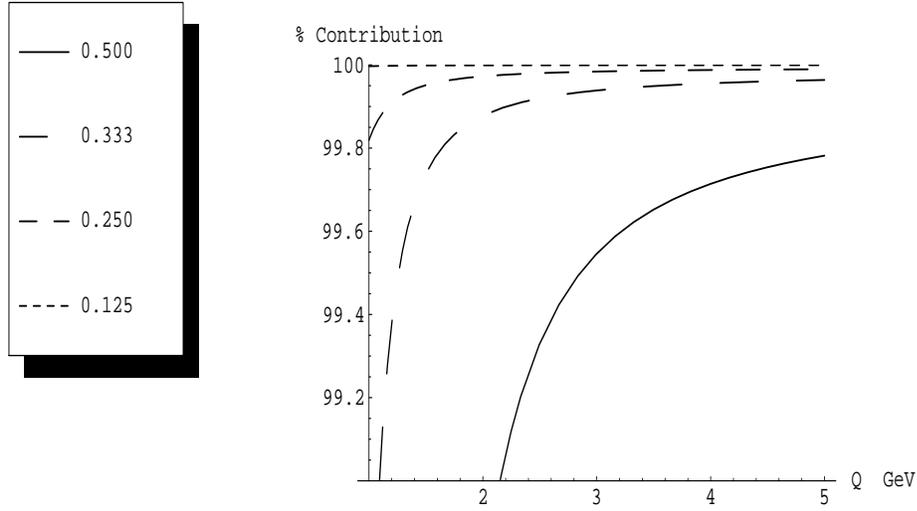,height=3in,width=5in}
\caption{Percent contribution to the form factor (Eq. \ref{e:hff}) from the region $0 \leq x \leq 1 - \lambda \frac{m}{Q}$ as a 
function of $Q$ in GeV for different $\lambda$. As indicated, the end-point region is \emph{not} dominant for asymptotic 
$Q$.}
\label{f:percent}
\end{center}
\end{figure}

In ascertaining the dominant contributions to the form factor in asymptopia, we have neglected the case $x=1$.
The form factor includes contributions from the end point, but this is where the scheme set up in Eq. \ref{e:regions} breaks down. 
So the above approximation Eq. \ref{e:asymf} is really only valid for $x \leq 1 - \lambda \frac{m}{Q}$, where $\lambda$ 
is some dimensionless cutoff less then one. For $1 - \lambda \frac{m}{Q} \leq x \leq 1$, we must return to the full expression for the form factor 
to get the end-point contribution. To leading order, however, $(1-x) \qperp \approx \lambda m \qperp / Q$ in the end-point region and 
the contribution to the form factor reads
\begin{equation} \label{e:end}
F(Q^2)^{\text{EP}} \approx \int_{1 - \lambda \frac{m}{Q}}^{1} \frac{dx}{x (1-x)} \int d\pperp \; \psi^{*}(x, \pperp + \lambda m \hat{q}^{\perp})
\psi(x, \pperp), 
\end{equation}
where $\hat{q}^{\perp} = \qperp / Q$ is the direction of $\qperp$. Since this contribution to the form factor depends only on $Q^2$, 
we can rotate our coordinates about the $z$-axis to make $\qperp$ parallel to $\hat{x}$. The resulting functional form is 
entirely similar to the full form factor, enabling swift evaluation:
\begin{equation} 
F(Q^2)^{\text{EP}} = \int_{1 - \lambda \frac{m}{Q}}^{1} f (x, \lambda^2 m^2 ) \; dx 
\end{equation}  
with $f(x, k^2)$ given by Eq. \ref{e:f}. The leading-order contribution to the above integral in asymptopia is found by 
expanding the integrand about $x =1$ and integrating. The result is
\begin{equation}  \label{e:fep}
F(Q^2)^{\text{EP}} = 8 \pi N (b^2 - a^2)^2 \;  \frac{ \lambda^6 + 2 \lambda^4 - 8  \lambda^2 - 2 \lambda (\lambda^2 + 1)
\sqrt{\lambda^2 + 4 }}{Q^4(\lambda^2 +  1)^3}  \log \Bigg[ \frac{\sqrt{\lambda^2 + 4 } - \lambda}
{\lambda^3 +  3 \lambda +  (\lambda^2 + 1) \sqrt{\lambda^2 + 4 } }  \Bigg] 
\end{equation}
which agrees with the Drell-Yan-West relation. To quantitatively consider the contribution from the end-point region, 
we have plotted the percent contribution to the form factor from
$ 0 \leq x \leq 1 - \lambda \frac{m}{Q}$. We have chosen 
the value of $\lambda$ to be smaller than one. Figure \ref{f:percent} 
 shows that the bulk of the form factor does not come from the end-point region. 

\subsection{Divide and Conquer} \label{s:cr}
In the process of trying to deduce the asymptotic behavior, our model has become infinitely sensitive to the end-point 
region. From the exact Hulth\'en form factor Eq. \ref{e:hff}, however, we know the end-point region does not overwhelmingly 
contribute (as Figure \ref{f:percent} confirms). Our dilemma sounds familiar \cite{L} and our approach, not surprisingly, 
is a regularization cutoff.  

To start, let us just toss away the troublesome divergent part of Eq. \ref{e:asymf} by introducing the cutoff 
$\lambda$ into the $x$ integral
\begin{equation} \label{e:cutoff}
F(Q^2) \to 2 \int^{1 - \lambda \frac{m}{Q}}_{0} \frac{dx}{x(1-x)} \psi^{*}(x, (1-x) \qperp) \int d\pperp \; \psi(x, \pperp).  
\end{equation}
The cutoff integral above can be computed analytically.  Using the integrand of Eq. \ref{e:asymf2},  the result reads
\begin{equation} \label{e:first} 
\lim_{Q \to \infty} \; F(Q^2) \sim \frac{32 \pi N (b^2 - a^2)^2} {Q^4} \Big[ \chi (a, b, m) - \log \lambda
 +\log \frac{Q}{m} \Big], 
\end{equation}
where $\chi(a,b,m)$ is a rather complicated, page-long function independent of $\lambda$. For our parameters $a, b$ and $m$, we have 
$\chi = 1.3635$. Nonetheless, we have discovered behavior via regularization which differs from $1/Q^{4}$ in a model 
which knows nothing about ultraviolet divergences, renormalization, \emph{etc}. 

One would think that with Eq. \ref{e:first}, we have determined the asymptotic behavior of the form factor. 
Although we threw away the end-point region to arrive at the above expression, we know precisely its contribution 
for a given $\lambda$, \emph{cf} Eq. \ref{e:fep}. The question remains: have we found all contributions to 
$\mathcal{O}[1/Q^4]$? 

There are order $1/Q^6$ corrections to the integrand of equation \ref{e:asymf2}. Putting these in give a 
correction term:
\begin{equation}
\frac{- 16 \pi m^2 N (b^2 - a^2)}{Q^6} \int_{0}^{1 - \lambda \frac{m}{Q}} \frac{dx \; x^2 }{(1-x)^4} 
\Big[ 2 x (1-x) \big( a^2 + b^2 \big) + (2x-1)^2 m^2 \Big] \log \Bigg[ 
\frac{4 x (1-x) b^2 + (2x-1)^2 m^2}{4 x(1-x) a^2 + (2x-1)^2 m^2} \Bigg] \notag
\end{equation}
Evaluating this correction term to leading order gives: $\frac{-32 \pi N (b^2 - a^2)^2}{Q^4 \lambda^2}$.
Thus terms in the integrand of order $1/Q^6$ give a contribution of order $1/Q^4$ to the asymptotic form factor. We haven't 
exhausted all of the $1/Q^6$ corrections, however--- we originally took only the first term in the Taylor expansion of 
Eq. \ref{e:ff} about $\pperp = (1-x) \qperp$ and the next non-vanishing term gives contributions of order $1/Q^6$.  
Even if we were to collect all the $\mathcal{O}[1/Q^6]$ corrections to the  $x$-integrand, we would have only just begun. 
One can easily find terms in the integrand of order $1/Q^8$ which emerge from the regularized $x$-integral $1/Q^4$. 
In fact, the integrand's correction terms of any order contribute to leading order in asymptopia! 

\begin{figure}
\begin{center}
\epsfig{file=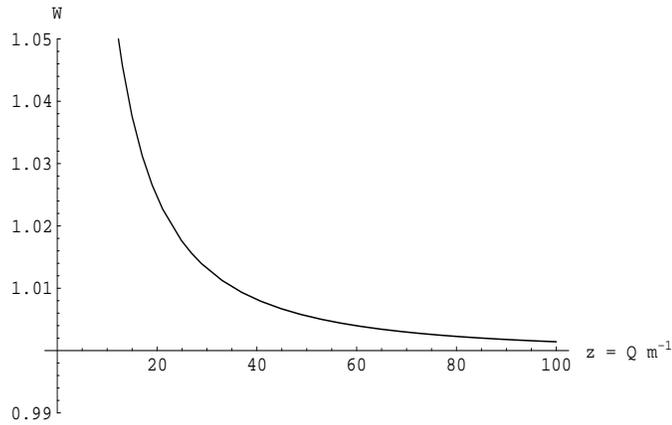,height=2.5in,width=3.5in}
\caption{Asymptotic Behavior of $F(Q^2)$ given by Equation \ref{e:hff}: $\rm{W} = \frac{z^4 F (z^2) - \alpha}{\beta \log z}$ 
is plotted as a function of $z = \frac{Q}{m}$ for the parameters $\alpha = 0.039472$ and $\beta = 0.044930$ 
which were graphically determined for $z$ around $400$.}
\label{f:finally}
\end{center}
\end{figure}

\begin{figure}
\begin{center}
\epsfig{file=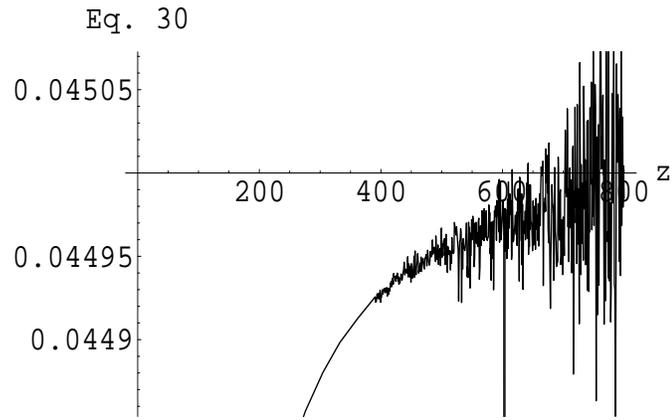,height=2.5in,width=3.5in}
\caption{Simple numerical determination of $\beta$ (via Equation \ref{e:beta}) plotted as a function $z = \frac{Q}{m}$. The numerical
integration clearly becomes imprecise past $z = 600$.}
\label{f:beta}
\end{center}
\end{figure}

Certainly we cannot hope to evaluate infinitely many leading-order terms. At least we have stumbled on to a prediction 
for the functional form in the asymptotic limit. Namely, we have seen 
\begin{equation}  \label{e:ab}
\lim_{z \to \infty} \; z^4 F(z^2) = \alpha + \beta \log z, 
\end{equation} 
with $z \equiv Q/m$.
We can test this prediction against the 
actual form factor's asymptotic limit calculated from equation \ref{e:hff}. In figure \ref{f:finally}, 
we test this hypothesis for empirically determined coefficients $\alpha = 0.039472$ and 
$\beta = 0.044930$  (calculated for $z$ around $400$). 
As the figure shows, this is indeed the form factor's behavior in asymptopia. 
It is quite curious to note: using Eq. \ref{e:first}, we would predict $\beta = 0.045020$ a difference of only $0.20 \%$
when compared with the empirical value. We believe this discrepancy results from approximating asymptotic $z$ as 
around $400$, not from ignoring infinitely many leading-order corrections. 
Indeed, we never found corrections of order $\log z/ z^4$ above, only a myriad of $1/z^4$ terms. 
It is our belief that the coefficient $\beta$ can be ascertained 
from the regularization integral Eq. \ref{e:first}.  The leading correction\footnote{Expanding the integrand of the form-factor
in powers of $Q$, there are only even terms. Once we exit the cutoff integral, however, we can now have any power of $Q$ in 
the expansion.} to $\beta$ determined graphically is $\mathcal{O}[1/ z^5]$---which 
gives a relative correction of $0.25 \%$ for $z \sim 400$. Taking $z$ larger in 
order to reduce this term only results in appreciable error in numerical integration. To verify our conjecture, we
have attempted to find the coefficient $\beta$ by varying $z$ . Figure \ref{f:beta} shows a plot of the graphically found
value of $\beta$ as a function of $z$ (the midpoint of our interval). Specifically we use a simple linear fit in the plot
\begin{equation}  \label{e:beta}
\beta (z) = \Bigg( (z+50)^4 F(z + 50) - (z - 50)^4 F(z- 50) \Bigg)  \log \Bigg[\frac{z - 50}{z + 50} \Bigg]. 
\end{equation}
The plot shows our cited value $\beta = 0.044930$  at $z = 400$. 
The trend is clear, $\beta$ is increasing to some limiting
value as $z$ increases. The numerical integration, however, becomes unreliable to $\sim 1\%$ past $600$. Nonetheless, it
appears we can determine $\beta$ from the regularization integral \ref{e:first}.  We are at a loss, however, 
to predict $\alpha$: there are simply an infinite number of correction terms to $\mathcal{O}[1/Q^4]$ to evaluate.  
 
\section{Similar Problems}
   \label{s:sim}
The problems encountered above are not unique to the Huth\'en model. In this section, we begin by exploring another model with 
similar behavior in asymptopia.  

\subsection{Coulomb Potential}
  \label{s:coulomb}
Let us suppose our particles interact via a Coulomb potential $V(r) \sim 1/r$, 
for which we take $ \psi(\mathbf{r}) = e^{-\kappa r}$. Then our momentum-space solution  to Eq. \ref{e:schr} is given by
\begin{align}  \label{e:wc}
\psi(x, \pperp) & \equiv\frac{m^3 \sqrt{N}}{16 (\kappa^2 + \mathbf{p}^2)^2} \notag \\
                & = \frac{ m^3 \sqrt{N} x^2 (1-x)^2}{(4 x (1-x) \kappa^2 + \pperp^2 + (2x-1)^2 m^2)^2}, 
\end{align}
where we have used Eq. \ref{e:p3} to re-express $p^{3}$ in the light-front center of mass. To find the asymptotic behavior 
of this
model's form factor, we use Eq. \ref{e:asymf}. Performing the integration, leaves us with a logarithmically divergent 
$x$-integral:
\begin{equation}
\lim_{Q \to \infty} \; F(Q^2) = \frac{m^6 N \pi}{Q^4} \int_{0}^{1} \frac{dx \; x^3}{(1-x) [ 4 x (1-x) \kappa^2 + (2x-1)^2 m^2]}.
\end{equation}   
Restricting $x$ to the range $0 \leq x \leq 1 - \lambda \frac{m}{ Q}$, and performing the integral yields
\begin{equation}
  \label{e:asymcoul}
\lim_{Q \to \infty} \; F(Q^2) = \frac{m^4 N \pi}{Q^4}  \Bigg[ \chi(\kappa, m) - \log \lambda  + \log  \frac{Q}{m}  \Bigg].
\end{equation}

\subsection{Why Regularization?}
As we have seen above, determining the exact asymptotic behavior of light-front form factors is no trivial task. 
The Drell-Yan-West relation is still apt at describing the contribution from the end-point region, however, this 
region does not dominate our form factors for asymptotic $Q$. Furthermore, techniques to determine the 
asymptotic behavior (brute force Taylor expansion, finding contributions from regions of minimal wave function 
suppression) led to logarithmically divergent $x$-integrals, suggesting the non-commutativity of the limits: $Q \to 
\infty$ and $x \to 1$. Here we show how      the potentials we use cause the
peculiarities for  $x \to 1$.

Starting with the expression for the light-front form factor Eq. \ref{e:ff}, we were led to the dominant contribution 
in asymptopia via isolating the regions of minimal wave function suppression, namely
\begin{equation} 
\lim_{Q \to \infty} \; F(Q^2) \approx 2 \int \frac{dx}{x (1-x)} \psi(x, (1-x) \qperp) \int d\pperp \psi (x, \pperp).\notag
\end{equation} 
Now let us utilize the Weinberg equation \cite{W}, which is the momentum-space version of our Eq. \ref{e:lfwf2}:
\begin{equation} \label{e:wine}
\psi(x, \pperp) = \frac{m^2/ 2 \pi^3}{M^2 - \frac{\pperp^2 + m^2}{x(1-x)}} \int \frac{dy \; d\mathbf{k}^
{\perp}}{y (1-y)} \psi(y, \mathbf{k}^{\perp}) V(x, \pperp; y, \mathbf{k}^{\perp}), 
\end{equation}
with $V$ as the Fourier transform of the potential. We can use this information in 
the asymptotic limit of the form factor, namely for
\begin{equation}
\psi(x, (1-x) \qperp) \approx  - \frac{m^2 x}{2 \pi^3 (1-x) Q^2} \int \frac{dy \; d\mathbf{k}^{\perp}}{y (1-y)}  
\psi(y, \mathbf{k}^{\perp}) V(x,(1-x) \qperp; y, \mathbf{k}^{\perp})
\end{equation}

Of course, in asymptopia the integral containing $V$ can be simplified. The potentials considered above are of the form
$V = V(\; | \mathbf{p}- \mathbf{k} | \;)$, where
\begin{equation} \label{e:diffp}
(\mathbf{p} - \mathbf{k})^2 = \Bigg( \pperp - \mathbf{k}^{\perp} \Bigg)^2 + \Bigg( (x - 1/2) 
\sqrt{\frac{\pperp^2 + m^2}{x (1-x)}} - (y - 1/2) 
\sqrt{\frac{\mathbf{k}^{\perp}{}^2 + m^2}{y (1-y)}} \Bigg)^2 
\end{equation}
which makes explicit use of having two equally massive particles in the center of mass frame (\emph{cf} Eq. \ref{e:p3}). 
As a result of the above equation, $V(x, (1-x)\qperp; y, \mathbf{k}^{\perp}) = V(x, (1-x) \qperp; 1/2, 0)$ 
to leading order. Henceforth we shall abbreviate $V(x, (1-x) \qperp; 1/2, 0)= V(\;|\mathbf{q}(x)|\;)$, where
\begin{equation}
  \label{e:q}
| \mathbf{q}(x) | = \sqrt{(1-x)^2 Q^2 + (q^3)^2} = \frac{Q}{2} \sqrt{\frac{1-x}{x}} 
\end{equation}
Revising the expression for the asymptotic form factor, we find (neglecting overall constants)
\begin{equation}  \label{e:it}
\lim_{Q \to \infty} \; F(Q^2) \sim \int  \frac{dx \; \phi(x)}{(1-x)^2 Q^2} V(\;|\mathbf{q}(x)|\;),
\end{equation}
where $\phi(x) = \int d\pperp \psi(x, \pperp)$ and the $y$-dependence integrated itself away. Eq. \ref{e:it} 
contains the answer to our troubled journey through asymptopia. At first glance, the integrand appears singular 
at the end point: containing one factor of $1-x$ from the measure, and another from the $(1-x)\qperp$ contained 
in the form factor. 
These factors are quite general and contain nothing specific about the interaction. While Eq. \ref{e:wellbehaved} 
spells out the criteria for good wave functions, it is necessary to be further 
restrictive by requiring $\psi(x, \pperp) \sim (1-x)^2$ as $x \to 1$ if we wish to cancel the potentially singular
denominator in Eq. \ref{e:it}. This isn't much of an imposition; both the Hulth\'en wave function Eq. \ref{e:wfp} and the 
Coulomb wave function Eq. \ref{e:wc} go like $(1-x)^2$ as $x \to 1$!

Only one $x$-dependent piece of Eq. \ref{e:it} remains to be considered---the potential. It is now immediately obvious that the 
Coulombic form factor should suffer logarithmic divergences in asymptopia:
\begin{equation}
V^{\text{C}}(\; |\mathbf{q}(x) | \; ) \sim - \frac{1}{\mathbf{q}(x)^2} =  - \frac{4 x}{(1-x) Q^2}.
\end{equation}
The potential brings along the anticipated factor of $Q^2$, but on the light front, an unwanted $(1-x)$ tags along. 
Given the behavior of the light front wave function, this extra factor is just enough to make the integral in Eq. \ref{e:it} 
diverge. The same is true for the Hulth\'en potential Eq. \ref{e:VH}, since we have
\begin{equation}
\lim_{Q \to \infty} V^{\text{H}}( \; |\mathbf{q}(x) | \;) = - \frac{12 \pi (a+ b)}
{\mathbf{q}(x)^2} = - \frac{48 \pi (a+ b) x}{(1-x) Q^2}.
\end{equation}

As illustrated above (Figures 1-4), the form factor itself isn't singular at asymptotic $Q$, 
just our means of obtaining it. This is made obvious by commuting the limits. Above we looked 
at $Q \to \infty$ first and found problematic behavior for  $x \to 1$ stemming from the potential. 
On the other hand,  consider taking $x$ near one first. 
We already did this for the Hulth\'en form factor, see Eq. \ref{e:end}. Now we are in a position to generalize 
this result. Since we know our wave functions $\psi(x, \pperp) \sim (1-x)^2$ as $x \to 1$, 
the contribution from the end point becomes:
\begin{equation} 
F(Q^2)^{\text{EP}} \sim \int_{1 - \lambda \frac{m}{Q}}^{1} (1-x)^3 dx  \to \frac{1}{Q^4}. \notag
\end{equation} 
This is just the Drell-Yan-West relation, which, as we have seen, does not account for the majority of the form factor 
in asymptopia. Clearly different behavior is seen when looking at asymptotic $Q$ near the end point, versus
the end-point region for asymptotic $Q$!
 
Logarithmically divergent form factors in asymptopia need not plague us any longer. The culprit has been unmasked: 
potentials in the \emph{auxiliary} coordinate-space,  such as  the $1/r$ of 
the Coulomb model or Eq. \ref{e:VH} for the Hulth\'en, which are  $\sim 1/r$ for small $r$ will lead to 
logarithmic divergences in the expression for the asymptotic
form factor. Of course, the asymptotic form factor 
\emph{itself}
is not singular. The logarithmic divergence of our asymptotic expression is a thorn-like warning: 
the series expansion in $1/Q$ does not converge uniformly in $x$.  

\subsection{Realistic Models on the Light Front}
Above we have seen rather simplistic models lead to electromagnetic form factors with non-standard asymptotic behavior. It is not likely, however, that this behavior is physical---though certainly it is the true asymptotic behavior for the models considered. 

Early work on factorization in quantum chromodynamics showed similar logarithms
appearing in asymptopia \cite{DM} for the nucleon electric form factor, and
thus a failure of renormalization group techniques.  Quite soon there after, it
was realized \cite{LB} that these logarithms were just 
manifestations of neglected higher-order corrections. Apparent end-point
singularities
are removed when evolution of the longitudinal momentum amplitude $\phi$  is
properly included, and consequently factorization is saved \cite{curious}.

Obviously our models do not have such higher-order corrections, and thus the
logarithm remains.
We must then wonder whether factorization breaks down for more realistic models
on the light front.
Let us then consider a more sophisticated model of the deuteron arising from
meson-theoretic potentials \cite{MA}. The general parameterization of the s-wave deuteron wave function is
\begin{equation}
\psi(r) = \frac{1}{r} \sum_{j = 1}^{N}  C_{j} e^{-m_{j} r}
\end{equation}   
with 
\begin{equation}
m_{j} = \alpha + (j - 1) m_{o}. \notag
\end{equation}
The usual boundary condition (finite wave function at the origin) leads to the constraint
\begin{equation} \label{e:sum}
\sum_{j = 1}^{N} C_{j} = 0. 
\end{equation} 

To put this realistic deuteron model on the light front, we work in momentum space and use the longitudinal momentum prescription above (Eq. \ref{e:p3}). The resulting form factor is completely similar to that of the Hulth\'en (Eq. \ref{e:hff}) except there are now $N^2$ terms instead of $4$. At this point, the similarity leads us to suspect $Q^4$ behavior modified by a logarithm in asymptopia. Based on our above analysis, verification of the logarithm's presence requires that we check $\psi(x, \pperp) \sim (1-x)^2$ as $x \to 1$ and that
the potential in $r$-space goes like $1/r$ near the origin. 

The momentum-space wave function has the end-point behavior
\begin{equation}
\psi(x, \pperp) \sim  \sum_{j = 1}^{N} C_{j} m_{j}^{2} \frac{(1-x)^2}{(m^2 + \pperp^2)^2} + \mathcal{O} \big[ (1-x)^3 \big],
\end{equation}
where the term linear in $1-x$ has vanished due to the constraint equation \ref{e:sum}. Appealing to the wave equation \ref{e:schr}, we can determine the potential which generates this deuteron wave function
\begin{equation}
V^{\text{MT}}(r) = \frac{\sum_{j = 1}^{N} C_{j} \Big[ 2 ( \alpha - m_{o}) m_{o} j + m_{o}^2 j^2 \Big] e^{- m_{o} r j }}{\sum_{\ell = 1}^{N} C_{\ell} e^{-m_{o} r \ell}} + \text{const}, 
\end{equation}
where the constant ensures the potential vanishes as $r \to \infty$. It is then straight forward to find the behavior near the origin
\begin{equation}
\lim_{r \to 0} V^{\text{MT}}(r) = - \frac{1}{r} \Big[ 2  (\alpha - m_{o}) + m_{o} \frac{\ell(N)}{k(N)} \Big], 
\end{equation}
where $\ell(N) = \sum j C_{j}$ and $k(N) = \sum j^2 C_{j}$. Thus based on these analytic observations, equation \ref{e:it} shows that even a realistic 
deuteron model will be troublesome in asymptopia.

\section{Conquering Asymptopia} \label{s:ca}
As we have seen, equation \ref{e:it} is a rather na\"{\i}ve way to determine a form factor's asymptotic behavior. This equation and the analysis leading 
to it were extrapolated from our knowledge of non-relativistic wave functions. Indeed one may verify that equation \ref{e:asymf} (derived by an analysis parallel to the non-relativistic one \cite{BL}) gives exactly the same results as equation \ref{e:it}. The breakdown of factorization for the above models is clearly 
a \emph{relativistic} problem (further verified by equation \ref{e:q}). Equation \ref{e:it} may not be useful in determining the asymptotic 
behavior. Without it, however, we wouldn't be aware of the cause of our problems at high momentum transfer. Now knowing when to expect 
trouble in asymptopia, let us proceed to correctly deduce the asymptotic behavior. 

\subsection{Wick-Cutkosky Model}
Before returning to the asymptotics of the Hulth\'en form factor, let us take a worthwhile look at the Wick-Cutkosky model. 
Consider two equally massive scalar particles
which interact by exchanging a massless scalar particle. The potential for such a process has been 
found and consequently
the ground-state wave function can be deduced using the momentum-space version
of equation \ref{e:schr}.
The wave function is
\cite{V1}
\begin{align}  \label{e:wcwf}
\psi(x, \pperp) & = \frac{8 \sqrt{\pi} \kappa^{5/2}}{(\mathbf{p}^2 + \kappa^2)^2 \Big( 1 + \big| \frac{p^3}{E(\mathbf{p})} \big| \Big)} \notag \\
                & = \frac{2^7 \sqrt{\pi} \kappa^{5/2} x^2 (1-x)^2 }{(4 x(1-x)\kappa^2 + \pperp^2 + (2 x -1)^2 m^2 )^2 ( 1+ |2 x - 1| )},
\end{align}
where we have used the energy $E(\mathbf{p}) = \frac{1}{2}\sqrt{\frac{\pperp^2 + m^2}{x (1-x)}}$ in the two-particle center of mass, 
and $p^3$ is given by equation \ref{e:p3}. Here $\kappa = 
\frac{1}{2} m a$ and $a = g^2 / 16 \pi m^2$ with $g$ as the coupling constant present in the interaction term. The invariant mass
of the system is $M = 2 m - \frac{1}{4} m a^2$.
To write this wave function, we have converted the explicitly covariant form cited in \cite{V1} into our own $z$-axis dependent form. The 
difference between these approaches does not concern us for the ground state of scalar particles. For a review of explicitly covariant
light-front dynamics see \cite{CDKM} or \cite{V4}. The wave function in equation
\ref{e:wcwf}
is  quite similar to our 
Coulomb wave function in section \ref{s:coulomb}. The
extra term in the denominator originates from retardation effects contained 
in the relativistic potential (effects which our na\"{\i}ve models clearly
lack). The behavior of the wave
function at the end point is not 
modified (to leading order in $1-x$) by retardation. Furthermore, the retarded potential is \cite{V1}
\begin{equation}
V(x,\pperp; y,\kperp)^{\text{WC}} = - 4 \pi a / \mathbf{\mathcal{K}}^2,
\end{equation}  
with
\begin{equation} 
\mathbf{\mathcal{K}}^2 = (\mathbf{p} - \mathbf{k})^2 - (2 x - 1)(2 y - 1)(E(\mathbf{p}) - E(\mathbf{k}))^2 + 
2 |x - y| \Big( E(\mathbf{p})^2 + E(\mathbf{k})^2 - \frac{M^2}{2} \Big), \notag
\end{equation}
where $(\mathbf{p} - \mathbf{k})^2$ is given by equation \ref{e:diffp}. Using this potential for asymptotic $Q$, we note
\begin{equation}
V(x, (1-x) \qperp; 1/2, 0)^{\text{WC}} \approx - \frac{16 \pi a x}{(1-x) Q^2 ( 1+ | 2 x -1 |)}
\end{equation}
which is singular at $x = 1$. Given this and the wave function's end-point behavior, we once again appeal to 
equation \ref{e:it} and a logarithmically divergent $x$-integral confronts us in deducing the asymptotic 
behavior of the form factor. Our experience above leads us to believe the true asymptotic behavior is $Q^{-4}$ 
modified by a logarithm. This asymptotic behavior for the Wick-Cutkosky model
has been found
previously by Karmanov and Smirnov \cite{V2} 
by considering regions which dominate the $x$ and $\pmag$ integrals of the form
factor.
The same asymptotic 
behavior of the Wick-Cutkosky model was also found  \cite{V2} by
using the Bethe-Salpeter approach. Karmanov and Smirnov state that 
 the logarithmic $\log{Q^2}/Q^4$  behavior
 was also been found earlier in \cite{AS}---a paper which admits 
the possibility of such logarithms only by announcing it is not considering
such cases. The presence of logarithmic modifications to relativistic form factors is discussed in \cite{GBB}, where the
authors interpret the Drell-Yan-West relation as valid modulo logarithms. 
Nonetheless, the correct asymptotic behavior
of the Wick-Cutkosky form factor was deduced in \cite{V2} as we shall now
demonstrate using techniques considered above. 

Considering the analytic form of the wave function \ref{e:wcwf}, we can see a region which dominates the form factor for asymptotic $Q$: 
$x$ near $\frac{1}{2}$. In this case, we can surely say $(1-x) Q \approx \frac{Q}{2} \gg m$ and consequently the analysis leading to equation \ref{e:asymf}
is certainly valid. Moreover, the Wick-Cutkosky wave function is identical to our Coulomb wave function for $x \approx \frac{1}{2}$.
Thus appealing to equation
\ref{e:asymf}, we find (to leading order about $x = \frac{1}{2}$)
\begin{align}
\lim_{Q \to \infty} \; F(Q^2) & = \frac{16 m^4 a^5}{\pi Q^4} \int_{0}^{1} \frac{dx}{\frac{a^2}{4} + (2 x - 1)^2} \notag \\
& = \frac{16 m^4 a^4}{Q^4}  \big( 1 + \mathcal{O}[a] \big),
\end{align}
where we have assumed $a \ll 1$ so that $x (1-x) a^2 \approx a^2/4$.
In order to compare with \cite{V2}, we have
been careful to adopt
their normalization (we have multiplied our expression \ref{e:ff} by $\frac{m}{2} (2 \pi)^{-3}$.) 

The other dominant contribution in asymptopia comes from \emph{near} the end-point region $\frac{1}{2} < (1-x) \ll 1 - \lambda \frac{m}{Q}$ 
as we have seen above by producing logarithms from regularized $x$-integrals. We can thus deduce the remaining contribution in asymptopia
via regularization. This result will be different from the Coulomb result, equation \ref{e:asymcoul}, due to the retardation factor. 
First we note that in the near end-point region $ x (1-x) a^2 \approx 0$. Now take $\frac{1}{2} 
< x_{o} < 1 - \lambda \frac{m }{Q}$ and hence the contribution which interests us reads  
\begin{align}
\lim_{Q \to \infty} \; F_{\lambda}(Q^2) & \equiv \frac{64 m^4 a^5}{\pi Q^4} \int_{x_{o}}^{1 - \lambda \frac{m}{Q}} \frac{x  \; dx}
{4 (1-x) (2 x -1)^2} \notag \\
    & = \frac{16 m^4 a^5}{\pi Q^4} \Bigg[ \xi(x_{o},a,m)  - \log \lambda + \log \frac{Q}{m} \Bigg] \big( 1 + \mathcal{O}\Big[\frac{1}{Q} 
\Big] \big)  
\end{align}  
Combining these two results (using only the logarithmic part of the latter), we arrive at the asymptotic behavior (to leading order in $a$)
\begin{equation} \label{e:asymwc}
\lim_{Q \to \infty} \; F(Q^2) \approx \frac{16 m^4 a^5}{\pi Q^4} \Big( \log \frac{Q}{m} + \frac{\pi}{a}   \Big)
\end{equation}
which agrees with the result found by Karmanov and Smirnov \cite{V2}. Furthermore one can use the Wick-Cutkosky wave function above to numerically
calculate the form factor and test equation \ref{e:asymwc} as predicting its asymptotic behavior. This form factor is less complicated
than the Hulth\'en model's. Consequently, the numerical integration is precise to larger $Q$. As before, 
we have graphically determined the coefficients $\alpha, \beta$ in equation \ref{e:ab} and observed the asymptotic behavior as 
tending toward $(\alpha + \beta \log z)/z^4$, with $z = \frac{Q}{m}$. The coefficient $\beta$
agrees well with equation \ref{e:asymwc}, differing by $< 0.1 \%$ for $a = 0.08, 0.007$. The error of the coefficient $\alpha$ depends
on how rapidly the series in $a$ converges. For $a = 0.08$, our graphically determined $\alpha$ differs from equation \ref{e:asymwc} 
by $\sim 12 \%$. While for $a = 0.007$, the error is $\sim 2 \%$. Indeed we have deduced the form factor's asymptotic behavior. 
  
\subsection{Back to the Hulth\'en}  
Our analysis above has been quite general and we shall now apply it to the Hulth\'en form factor. To finish our quest through asymptopia, 
it remains to determine the coefficient $\alpha$ in equation \ref{e:ab} for the Hulth\'en model. As we learned above\footnote{
A more precise statement
is: expanding about $x = \frac{1}{2}$ in equation \ref{e:asymf} gives the coefficient $\alpha$ up to a possible factor of $2$.}, considering 
the contribution for $x \approx \frac{1}{2}$ gives us $\alpha$ for asymptotic $Q$. So we return to equation \ref{e:asymf2} 
and expand to leading order about $x = \frac{1}{2}$
\begin{equation}
\lim_{Q \to \infty} \; F(Q^2) \approx \frac{4 \pi m^2 N (b^2 - a^2)}{Q^4} \int_{0}^{1} \log \Bigg[ \frac{b^2 + (2 x - 1)^2 m^2}
{a^2 + (2 x - 1)^2 m^2} \Bigg] dx. 
\end{equation} 
having used $\frac{a}{m}, \frac{b}{m} \ll 1$. At first glance, it appears we have dropped a factor of two from the asymptotic
expression \ref{e:asymf2}. However careful consideration of region i (in equation \ref{e:regions}), shows its contribution vanishes 
(to leading order in $1/Q$). Evaluating the above integral and combining with the 
logarithmic part of our previous result (equation \ref{e:first}), we find
\begin{equation} \label{e:theend}
\lim_{Q \to \infty} \; F(Q^2) = \frac{32 \pi N (b^2 -a^2)^2}{Q^4}  \Bigg[\log \frac{Q}{m} + \frac{1}{8} \Bigg( \frac{m \pi}{b + a} - 1 \Bigg) \Bigg].
\end{equation} 
From which we deduce $\alpha = 0.046493$  (and $\beta = 0.045020$ as discussed previously).
Comparing with the graphically determined result of section \ref{s:cr}, we see that there is
a $17.8 \%$ difference. Again this difference is due to the series expansion in small parameters: 
$a/m =0.048943 $ and $b/m = 0.29007$. For smaller values of the parameters, we expect better results. 
However, with smaller parameters one needs higher $Q$ to graphically determine $\alpha, \beta$ and the numerical integration 
becomes imprecise. Nonetheless, within our constraints we have verified equation \ref{e:theend} as the 
asymptotic behavior of the Hulth\'en form factor. 

\section{Concluding Remarks}
We have undertaken a relatively simple task
to compare relativistic and non-relativistic
form factors for the Hulth\'en model of the deuteron.
For small momentum transfer, the two versions differ by about a percent and the
root mean square radii differ by even less.
The behavior for large $Q$, however, led us on an unexpected journey. 

Our expedition through asymptopia helped us learn the Hulth\'en form
factor's
true behavior $\sim \log Q^2/ Q^4$ for large $Q$.
The path was circuitous  because
the conventional means
(asymptotic expressions, Taylor series expansions) lead directly
to logarithmic divergences. These difficulties are
manifestations of the non-commutativity of the limits $Q \to \infty$
and $x \to 1$, and hence indicative of the breakdown of factorization. 

Indeed, we find that  this behavior is not
particular to the Hulth\'en model. Equation \ref{e:it}
tracks down the root of these  difficulties.
Generating our light front wave functions from non-relativistic potentials
singular at the origin will lead to problematic relativistic form factors in
the asymptotic limit.
Such problems do not plague calculations in fundamental theories because
higher-order corrections necessarily cancel the divergences. For realistic
models,
however, the breakdown of factorization persists and is an obstruction to straightforward asymptotic 
calculations. 

\begin{center}
{\bf Acknowledgment}
\end{center}
This work was funded by the U.~S.~Department of Energy, grant: DE-FG$03-97$ER$41014$.


\begin{thebibliography}{99}
\bibitem{h123}        
Starting with a two body light-front Hamiltonian
$ 
        P^{-} \psi = \Big( \frac{\koneperp^{2} + m^2}{\koneplus} + 
        \frac{\ktwoperp^{2} + m^2}{\ktwoplus} + \mathcal{V} \Big) \psi  
$ 
and repeating the above algebra in the center of mass frame, we see that 
$V^{H} = M \mathcal{V} / 4$.
\bibitem{numb}  
There are small errors in numerical integration, however. 
They amount
to only one part in $10^{10}$ for the range of $Q$ plotted in Figure \ref{f:rnrhighQ}. We must go to 
higher $Q$ for the error to accumulate to $\sim 1\%$, see Figure \ref{f:beta}.
\bibitem{curious} It is curious to note that the functional forms involved for the light front
  Hulth\'en model are
  remarkably similar to end-point region suppression via Sudakov form factors
  due to gluon
  corrections to the quark-photon vertex (see Appendix E of \cite{LB}) with the
  exception that our transverse integrals are finite.
\end{thebibliography}
\end{document}